\documentclass[twocolumn,tighten]{aastex62}
\usepackage{color}

\usepackage{multirow}

\shortauthors{Sano et al.}

\received{June 27, 2018}
\revised{September 3, 2018}
\accepted{September 7, 2018}

\begin{document}
\vspace*{-1cm}
\title{Molecular Clouds associated with the Type Ia SNR N103B in the Large Magellanic Cloud}

\author{H. Sano}
\affiliation{Institute for Advanced Research, Nagoya University, Furo-cho, Chikusa-ku, Nagoya 464-8601, Japan; sano@a.phys.nagoya-u.ac.jp}
\affiliation{Department of Physics, Nagoya University, Furo-cho, Chikusa-ku, Nagoya 464-8601, Japan}

\author{Y. Yamane}
\affiliation{Department of Physics, Nagoya University, Furo-cho, Chikusa-ku, Nagoya 464-8601, Japan}

\author{K. Tokuda}
\affiliation{Department of Physical Science, Graduate School of Science, Osaka Prefecture University, 1-1 Gakuen-cho, Naka-ku, Sakai 599-8531, Japan}
\affiliation{National Astronomical Observatory of Japan, Mitaka, Tokyo 181-8588, Japan}

\author{K. Fujii}
\affiliation{Department of Astronomy, School of Science, The University of Tokyo, 7-3-1 Hongo, Bunkyo-ku, Tokyo 133-0033, Japan}

\author{K. Tsuge}
\affiliation{Department of Physics, Nagoya University, Furo-cho, Chikusa-ku, Nagoya 464-8601, Japan}

\author{T. Nagaya}
\affiliation{Department of Physics, Nagoya University, Furo-cho, Chikusa-ku, Nagoya 464-8601, Japan}

\author{S. Yoshiike}
\affiliation{Department of Physics, Nagoya University, Furo-cho, Chikusa-ku, Nagoya 464-8601, Japan}

\author{M. D. Filipovi{\'c}}
\affiliation{Western Sydney University, Locked Bag 1797, Penrith South DC, NSW 1797, Australia}

\author{R. Z. E. Alsaberi}
\affiliation{Western Sydney University, Locked Bag 1797, Penrith South DC, NSW 1797, Australia}

\author{L. Barnes}
\affiliation{Western Sydney University, Locked Bag 1797, Penrith South DC, NSW 1797, Australia}

\author{T. Onishi}
\affiliation{Department of Physical Science, Graduate School of Science, Osaka Prefecture University, 1-1 Gakuen-cho, Naka-ku, Sakai 599-8531, Japan}

\author{A. Kawamura}
\affiliation{National Astronomical Observatory of Japan, Mitaka, Tokyo 181-8588, Japan}

\author{T. Minamidani}
\affiliation{Nobeyama Radio Observatory, Minamimaki-mura, Minamisaku-gun, Nagano 384-1305, Japan}
\affiliation{Department of Astronomical Science, School of Physical Science, SOKENDAI (The Graduate University for Advanced Studies), 2-21-1, Osawa, Mitaka, Tokyo 181-8588, Japan}

\author{N. Mizuno}
\affiliation{Nobeyama Radio Observatory, Minamimaki-mura, Minamisaku-gun, Nagano 384-1305, Japan}

\author{H. Yamamoto}
\affiliation{Department of Physics, Nagoya University, Furo-cho, Chikusa-ku, Nagoya 464-8601, Japan}

\author{K. Tachihara}
\affiliation{Department of Physics, Nagoya University, Furo-cho, Chikusa-ku, Nagoya 464-8601, Japan}

\author{N. Maxted}
\affiliation{Western Sydney University, Locked Bag 1797, Penrith South DC, NSW 1797, Australia}
\affiliation{School of Physics, The University of New South Wales, Sydney, 2052, Australia}

\author{F. Voisin}
\affiliation{School of Physical Sciences, The University of Adelaide, North Terrace, Adelaide, SA 5005, Australia}

\author{G. Rowell}
\affiliation{School of Physical Sciences, The University of Adelaide, North Terrace, Adelaide, SA 5005, Australia}

\author{H. Yamaguchi}
\affiliation{Institute of Space and Astronautical Science, JAXA, 3-1-1 Yoshinodai, Sagamihara, Kanagawa 229-8510, Japan}
%\affiliation{NASA Goddard Space Flight Center, Code 662, Greenbelt, MD 20771, USA}
%\affiliation{Department of Astronomy, University of Maryland, College Park, MD 20742, USA}

\author{Y. Fukui}
\affiliation{Institute for Advanced Research, Nagoya University, Furo-cho, Chikusa-ku, Nagoya 464-8601, Japan; sano@a.phys.nagoya-u.ac.jp}
\affiliation{Department of Physics, Nagoya University, Furo-cho, Chikusa-ku, Nagoya 464-8601, Japan}

\begin{abstract}
N103B is a Type Ia supernova remnant (SNR) in the Large Magellanic Cloud (LMC). We carried out new $^{12}$CO($J$ = 3--2) and $^{12}$CO($J$ = 1--0) observations using ASTE and ALMA. We have confirmed the existence of a giant molecular cloud (GMC) at $V_\mathrm{LSR}$ $\sim$245 km s$^{-1}$ towards the southeast of the SNR using ASTE $^{12}$CO($J$ = 3--2) data at an angular resolution of $\sim$25$\arcsec$ ($\sim$6 pc in the LMC). Using the ALMA $^{12}$CO($J$ = 1--0) data, we have spatially resolved CO clouds along the southeastern edge of the SNR with an angular resolution of $\sim$1.8$\arcsec$ ($\sim$0.4 pc in the LMC). The molecular clouds show an expanding gas motion in the position--velocity diagram with an expansion velocity of $\sim5$ km s$^{-1}$. The spatial extent of the expanding shell is roughly similar to that of the SNR. We also find tiny molecular clumps in the directions of optical nebula knots. We present a possible scenario that N103B exploded in the wind-bubble formed by the accretion winds from the progenitor system, and is now interacting with the dense gas wall. This is consistent with a single-degenerate scenario.
\end{abstract}


\section{Introduction} \label{sec:intro}
Identifying the progenitor system of Type Ia supernovae is a long-standing problem of modern astrophysics, especially in light of their use as standard candles in cosmology \citep[e.g.,][]{1999ApJ...517..565P}. At present, two theoretical scenarios have been proposed: A single-degenerate (SD) scenario in which a white dwarf has accreted materials from a non-degenerate companion---a red giant, main-sequence star, or a helium star---until the white dwarf mass gets close to the Chandrasekhar limit $\sim$1.4 $M_\sun$ \citep{1973ApJ...186.1007W,1982ApJ...257..780N,1984ApJS...54..335I,1985ASSL..113....1P}, and a double-degenerate (DD) scenario represents the merger of two white dwarfs \citep{1982ApJ...257..780N,1984ApJ...277..355W}.

To distinguish these scenarios, The identification of the companion star and the associated dense circumstellar medium (CSM) is thought to be essential. For the SD scenario, the non-degenerate companion can survive after the explosion of the primary white dwarf \citep[e.g.,][]{2010ApJ...708.1025K}, and binary interactions may produce the dense CSM whose chemical composition is modified by the mass loss from the progenitor system \citep{2008ApJ...679.1390H}. For the DD scenario, no companion star or dense CSM are expected. Pioneering studies have succeeded in finding the ionized CSM in Kepler's supernova remnant (SNR) \citep[][]{2001A&A...373..281D,2007ApJ...662..998B,2012ApJ...755....3W}. The CSM in Kepler's SNR shows N overabundance, which would be originated by CNO-processed medium from the progenitor system, providing a further support for the SD origin \citep{2015ApJ...808...49K}.

Investigating the interstellar neutral gas around Type Ia SNRs has received attention as an alternative approach to understand the nature of progenitor systems \cite{2016ApJ...826...34Z} discovered an expanding bubble surrounding Tycho's SNR based on the CO data obtained with the IRAM 30 m telescope. The expanding velocity of the bubble was found to be $\sim$4.5 km s$^{-1}$ and has a mass of $\sim$220 $M_\sun$. Enhanced intensity ratio of $^{12}$CO $J$ = 2--1 / 1--0 and possible line broadening were also found at the northeastern rim of the SNR. The authors concluded that the bubble was driven by the accretion wind with the velocity of a few hundreds km s$^{-1}$, suggesting the SD scenario as the progenitor system of Tycho's SNR. Subsequently, \cite{2017A&A...604A..13C} found double shell-like structures of CO clouds around Tycho's SNR. The large CO cavity---a size of 13 pc $\times$ 27 pc---is partially connected with the small CO cavity with a size of $\sim$9 pc, both of which are expanding at a velocity of $\sim3$--4 km s$^{-1}$. The total kinematic energy of CO cavities is consistent with the prediction of the accretion wind model \citep[e.g.,][]{1996ApJ...470L..97H,1999ApJ...522..487H,1999ApJ...519..314H}. The authors therefore concluded that the large cavity cloud be also formed by the accretion winds from the SD progenitor system of the SNR. To better understand the SD scenario, we need more samples of the Type Ia SNRs associated with the interstellar neutral gas.

N103B (also known as SNR 0509$-$68.7) is one of the youngest Type Ia SNRs in the Large Magellanic Cloud (LMC). The light echo observations estimated the age to be 380--860 yr \citep{2005Natur.438.1132R}. The angular size of N103B is $27'' \times 29''$ corresponding the diameter $\sim$6.8 pc at the LMC distance of 50 kpc \citep[e.g.,][]{2016A&A...585A.162M,2017ApJS..230....2B}. The {\it{Chandra}} X-ray image (Figure \ref{fig1}a) unveils the bright western-half comprising a series of dense knots and diffuse emission \citep{2003ApJ...582..770L}. Studies of the X-ray spectroscopy show the large mass of Fe ($\sim$0.34 $M_\sun$) and the lack of a region of O-rich ejecta, indicating a Type Ia origin for the SNR \citep{1995ApJ...444L..81H,2003ApJ...582..770L}. The presence of highly-ionized Fe suggests the strong interaction between the supernova ejecta and the dense CSM \citep{2014ApJ...785L..27Y}. In fact, recent optical and infrared photometric/spectroscopic observations identified the ionized CSM with a density of $\sim$500--5000 cm$^{-3}$ and a possible candidate for a surviving main-sequence companion near the center of the SNR, suggesting a likely candidate for the SD origin \citep{2017ApJ...836...85L, 2017ApJ...847..122G}. The SD scenario is also supported by measurements of the Ca/S mass ratios by X-ray spectroscopic analysis \citep{2017ApJ...843...35M}.

In contrast, detailed studies of the interstellar neutral gas of N103B have not yet been reported. \cite{2017AIPC.1792d0038S,sano2018} revealed a giant molecular cloud (GMC) in the direction of the SNR N103B using the Mopra 22-m radio telescope. The authors estimated the total mass of GMC to be $\sim$$10^6$ $M_\sun$, which is significantly higher than the masses of GMCs associated with the LMC SNRs N132D and N49, and the LMC superbubble 30~Doradus~C \citep{1997ApJ...480..607B,2015ASPC..499..257S,2017ApJ...843...61S,2018ApJ...863...55Y}. Despite the promising candidate for the site of shock-cloud interaction, the authors could not resolve the spatial distribution of molecular clouds likely associated with the SNR N103B due to the limited angular resolution of Mopra ($\sim$45$''$ corresponding to $\sim$11 pc in the LMC). 

In the present paper, we report new millimeter/submillimeter wavelength observations using $^{12}$CO($J$ = 1--0, 3--2) emission lines with the Atacama Large Millimeter/submillimeter Array (ALMA) and the Atacama Submillimeter Telescope Experiment (ASTE). Because of the high angular resolution of  $\sim$1.8$''$ and $\sim$25$''$ ($\sim$0.4 pc and $\sim$6 pc in the LMC), we can spatially resolve the GMC into individual molecular clouds. Section \ref{sec:obs} gives observations and data reductions of CO, H{\sc i}, and X-rays. Subsection \ref{subsec:large} overviews the large-scale Distribution of the X-rays, CO, and H{\sc i}; subsections \ref{subsec:alma} and \ref{subsec:tiny} present a detailed CO distribution with ALMA. Discussion and conclusions are given in Sections \ref{sec:dis} and \ref{sec:conc}, respectively. In a subsequent paper (Alsaberi et al. 2018 in preparation), we will examine radio continuum component of the N103B with special emphasis on its expansion, polarization and spectral energy distribution.

\begin{deluxetable*}{cccccccc}[]
\tablewidth{\linewidth}
\tablecaption{Summary of $Chandra$ archive data}
\tablehead{
Observation ID &  $\alpha_{\mathrm{J2000}}$ & $\delta_{\mathrm{J2000}}$ & Start Date  & Exposure & Roll Angle \\
& ($^\mathrm{h}$ $^\mathrm{m}$ $^\mathrm{s}$) & ($\degr$ $\arcmin$ $\arcsec$) & (yyyy-mm-dd hh:mm:ss) & (ks) & ($\degr$) }
\startdata
\phantom{0}\phantom{0}125 & 05 08 59.00 & $-68$ 43 30.00 & 1999-12-04 12:26:57 & 36.7 & \phantom{0}$+71$\\
\phantom{0}3810 & 05 08 44.00 & $-68$ 45 36.00 & 2003-02-07 01:29:46 & 29.7 & $-152$\\
18018 & 05 08 59.00 & $-68$ 43 34.00 & 2017-03-20 18:12:45 & 39.5 & \phantom{0}$+47$\\
20042 & 05 08 59.00 & $-68$ 43 34.00 & 2017-03-24 18:59:37 & 19.8 & \phantom{0}$+39$\\
19921 & 05 08 59.00 & $-68$ 43 34.00 & 2017-04-10 22:26:08 & 16.8 & \phantom{0}$+19$\\
20058 & 05 08 59.00 & $-68$ 43 34.00 & 2017-04-14 13:57:32 & 43.8 & \phantom{0}$+20$\\
18020 & 05 08 59.00 & $-68$ 43 34.00 & 2017-04-17 21:23:32 & 27.2 & \phantom{0}$+25$\\
19923 & 05 08 59.00 & $-68$ 43 34.00 & 2017-04-26 19:06:00 & 58.3 & \phantom{0}$+26$\\
20067 & 05 08 59.00 & $-68$ 43 34.00 & 2017-05-04 18:49:53 & 29.7 & \phantom{0}$-20$\\
20053 & 05 08 59.00 & $-68$ 43 34.00 & 2017-05-22 11:42:30 & 11.2 & \phantom{0}$-61$\\
18019 & 05 08 59.00 & $-68$ 43 34.00 & 2017-05-23 19:08:22 & 59.3 & \phantom{0}$-37$\\
19922 & 05 08 59.00 & $-68$ 43 34.00 & 2017-05-30 01:23:59 & 41.4 & \phantom{0}$-48$\\
20085 & 05 08 59.00 & $-68$ 43 34.00 & 2017-05-31 04:07:23 & 14.9 & \phantom{0}$-54$\\
20074 & 05 08 59.00 & $-68$ 43 34.00 & 2017-06-01 19:08:17 & 31.2 & \phantom{0}$-55$\\
\enddata
\tablecomments{All exposure times represent the net available times alter several filtering procedures are applied in the data reduction processes.\vspace*{-0.5cm}}
\label{tab1}
\end{deluxetable*}

\section{Observations and Data Reductions}\label{sec:obs}
\subsection{CO}\label{subsec:co}
Observations of $^{12}$CO($J$ = 3--2) line emission at 0.87 mm wavelength were carried out in 16--24 November 2015 using ASTE \citep{2004SPIE.5489..763E}. We used the on-the-fly (OTF) mapping mode with Nyquist sampling. The map size is $2' \times 2'$ rectangular region with a center position at ($\alpha_\mathrm{J2000}$, $\delta_\mathrm{J2000}$) $\sim$ ($05^\mathrm{h}08^\mathrm{m}59^\mathrm{s}$, $-68\degr43\arcmin36\arcsec$). The side-band-separating mixer receiver ``DASH 345'' receiver was used for a front end. The back end was ``MAC'' digital FX spectrometer \citep{2000SPIE.4015...86S} with 1024 channels with a bandwidth of 128 MHz, corresponding to the velocity coverage of $\sim$111 km s$^{-1}$ and the velocity resolution of $\sim$0.11 km s$^{-1}$. The typical system temperature was $\sim$300--600 K in the single-side-band (SSB), including the atmosphere. We smoothed the data with a two-dimensional Gaussian function of $11''$. The final beam size is $\sim$25$''$ (FWHM). The pointing accuracy was checked every half-hour to satisfy an offset within $2''$. We also observed N159W [($\alpha_\mathrm{J2000}$, $\delta_\mathrm{J2000}$) $\sim$ ($05^\mathrm{h}40^\mathrm{m}3\fs7$, $-68\degr47\arcmin00\arcsec$)] \citep{2011AJ....141...73M} for the absolute intensity calibration, and then we estimated the main beam efficiency of $\sim$0.55. The final data has a noise fluctuation of $\sim$0.11 K at the velocity resolution of $\sim$0.4 km s$^{-1}$.

Observations of $^{12}$CO($J$ = 1--0) line emission at 2.6 mm wavelength were conducted at 31th January and 27th August 2016 using ALMA Band 3 (84--119 GHz) as a Cycle 3 filler project $\#$2015.1.01130.S (PI: Kosuke Fujii). We used the mosaic mapping mode of a $100'' \times 100''$ rectangular region with a center position at ($\alpha_\mathrm{J2000}$, $\delta_\mathrm{J2000}$) $\sim$ ($05^\mathrm{h}08^\mathrm{m}59^\mathrm{s}$, $-68\degr43\arcmin36\arcsec$). The observations were carried out using 38 antennas for 12 m array. The baseline ranges are from 13.7 to 1551.1 m, corresponding to the {\it{u-v}} coverage form 4.6 to 596.0 $k\lambda$. The correlator was set to have a bandwidth of 58.59 MHz, representing the velocity coverage of 152.5 km s$^{-1}$. The calibration used three quasars: J0635$-$7516 for the complex gain calibrator, J0519$-$4546 for the flux calibrator, and J0529$-$7245 for the phase calibrator. These calibrations and data reduction were made by the Common Astronomy Software Application \citep[CASA;][]{2007ASPC..376..127M} package version 5.1.0. We used the natural weighting and the multiscale CLEAN algorithm \citep{2008ISTSP...2..793C} implemented in the CASA package. The final beam size is $1.92'' \times 1.73''$ with a position angle of $66\fdg7$, corresponding to the spatial resolution of $\sim$0.44 pc at the LMC distance of 50 kpc. Typical noise fluctuation is $\sim$0.95 K at 0.4 km s$^{-1}$ velocity resolution. To estimate the missing flux, we compared the integrated intensity of the Mopra single-dish data \citep{2011ApJS..197...16W,2017ApJ...850..139W,2017AIPC.1792d0038S,sano2018} with the ALMA CO data that smoothed to match the FWHM of the single-dish data ($\sim$45$''$). Since the total integrated intensity of ALMA data is $\sim$10$\%$ smaller than that of the single-dish data, the missing flux is considered to be negligible.

\begin{figure*}
\begin{center}
\includegraphics[width=\linewidth,clip]{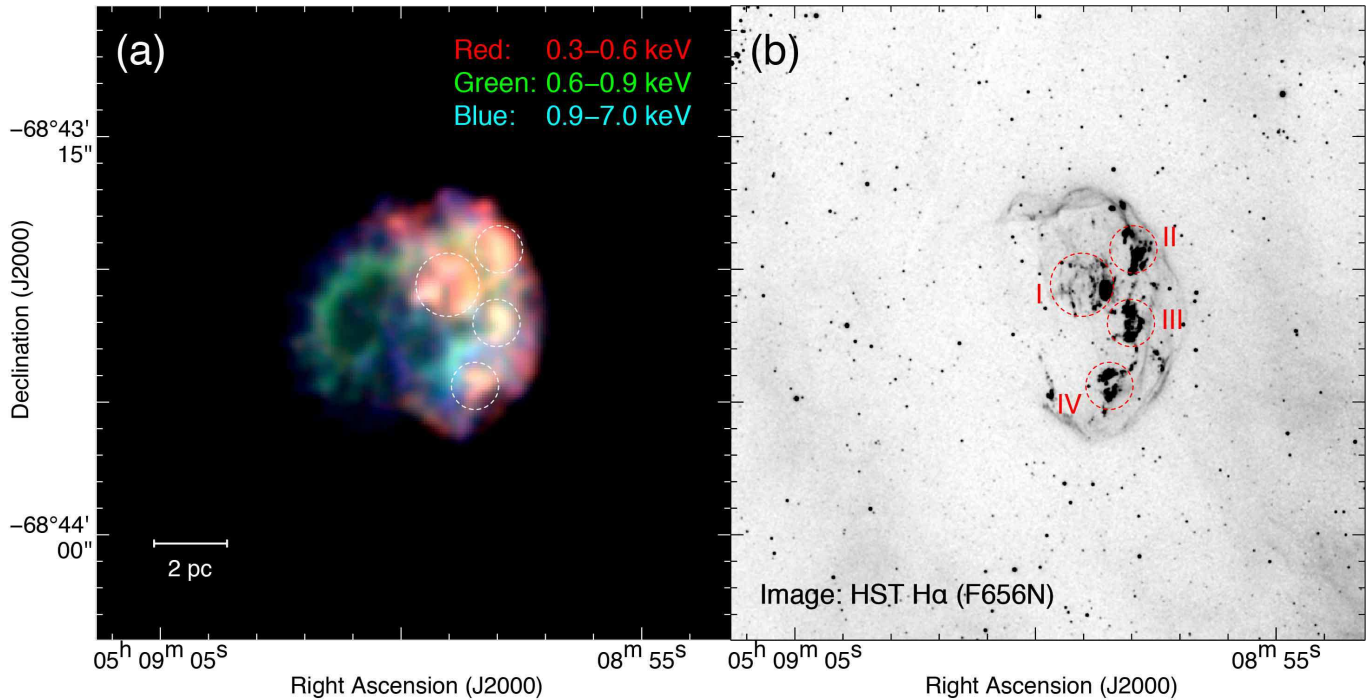}
\caption{(a) RGB image of X-rays obtained with Chandra \citep{2003ApJ...582..770L}. The red, green, and blue represent the energy bands, 0.3--0.6 keV, 0.6--0.9 keV, and 0.9--7.0 keV, respectively. (b) H$\alpha$ image using {\it{Hubble Space Telescope}} (HST) with the FW656N filter \citep{2017ApJ...836...85L}. Dashed circles indicate the optical nebular knots (I), (II), (III), and (IV) as defined by \citep{2017ApJ...836...85L}. We also annotate the dashed circles in Figure \ref{fig1}a.}
\label{fig1}
\end{center}
\end{figure*}%

\subsection{H{\sc i}}\label{subsec:hi}
We also used archival H{\sc i} 21 cm line data published by \cite{2003ApJS..148..473K}. The data were obtained using the Australia Telescope Compact Array (ATCA) and combined with single-dish data from Parkes radio telescope operated by Australia Telescope National Facility. The angular resolution of the H{\sc i} data is $60''$, corresponding to the spatial resolution of $\sim$15 pc at the distance of the LMC. The sensitivity of brightness temperature is $\sim$2.4 K for the velocity resolution of 1.689 km s$^{-1}$. 

\subsection{X-rays}\label{subsec:xray}
Fourteen {\it{Chandra}} pointed observation data with the Advanced CCD Imaging Spectrometer S-array (ACIS-S3) are available for N103B as summarized in Table \ref{tab1}. The data that the Obs IDs are 125 and 3810 have been published in \cite{2003ApJ...582..770L}. We used {\it{Chandra}} Interactive Analysis of Observations \cite[CIAO;][]{2006SPIE.6270E..1VF} software version 4.10 and CALDB 4.7.8 for data reduction and imaging analysis. The datasets was reprocessed using the ``chandra\_repro" procedure. We created combined, energy-filtered, exposure-corrected images using the ``merge\_obs" procedure in the energy bands of 0.3--0.6, 0.6--0.9, 0.9--7.0 keV, and 0.3--7.0 keV. The total effective exposure time and pixel size are 459.5 ks and $0\farcs49$, respectively.

\begin{figure*}
\begin{center}
\includegraphics[width=\linewidth,clip]{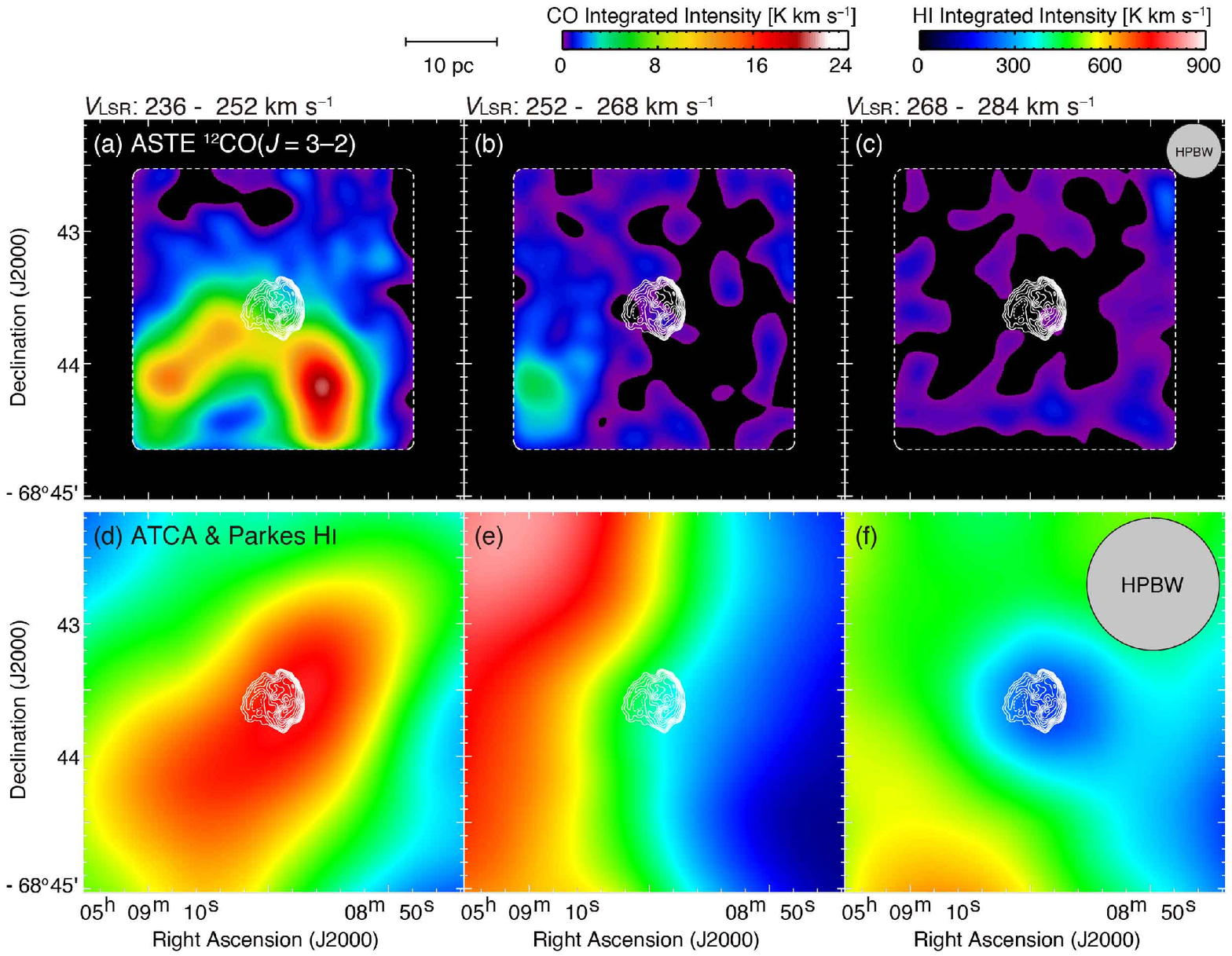}
\caption{Velocity channel maps of (a--c) ASTE $^{12}$CO($J$ = 3--2) and (d--f) ATCA $\&$ Parkes H{\sc i}. The integration velocity range is $V_\mathrm{LSR}$ = 236--252 km s$^{-1}$ for Figures \ref{fig2}(a) and \ref{fig2}(d), $V_\mathrm{LSR}$ = 252--268 km s$^{-1}$ for Figures \ref{fig2}(b) and \ref{fig2}(e), and $V_\mathrm{LSR}$ = 268--284 km s$^{-1}$ for Figures \ref{fig2}(c) and \ref{fig2}(f). Superposed contours represent {\it{Chandra}} X-rays in the energy band from 0.3--7.0 keV. The contour levels are 1.0, 1.5, 2.1, 2.8, 3.6, 4.5, 5.5, 6.6, 7.8, and $9.0 \times 10^{-6}$ photons s$^{-1}$ pixel$^{-1}$. The beam sizes of CO and H{\sc i} are shown in Figures  \ref{fig2}(c) and  \ref{fig2}(f), respectively. The dashed lines in Figures  \ref{fig2}(a--c) indicate the observed area of CO.}
\label{fig2}
\end{center}
\vspace*{0.1cm}
\end{figure*}%

\section{Results}\label{sec:results}
\subsection{Large-Scale Distribution of the X-rays, CO, and H{\sc i}}\label{subsec:large}
Figure \ref{fig1}a shows a three-color X-ray image of the LMC SNR N103B as observed by {\it{Chandra}} \citep[e.g.,][]{2003ApJ...582..770L}. The soft-band X-rays (red: 0.3--0.6 keV) are bright in the western shell, while the medium-band X-rays (green: 0.6--0.9 keV) and hard-band X-rays (blue: 0.9--7.0 keV) show a similar brightness in both the western and eastern half. The northwestern shell has a spherical edge in the soft-band X-rays, whereas the southeastern shell is strongly deformed especially toward the regions at ($\alpha_\mathrm{J2000}$, $\delta_\mathrm{J2000}$) $\sim$ ($05^\mathrm{h}09^\mathrm{m}00^\mathrm{s}$, $-68\degr43\arcmin43\arcsec$) and at $\sim$ ($05^\mathrm{h}08^\mathrm{m}58^\mathrm{s}$, $-68\degr43\arcmin46\arcsec$). The X-ray bright spots are also located not only in the shell interior around ($\alpha_\mathrm{J2000}$, $\delta_\mathrm{J2000}$) $\sim$ ($05^\mathrm{h}09^\mathrm{m}58\fs0$, $-68\degr43\arcmin33\arcsec$), but also in the western shell at ($\alpha_\mathrm{J2000}$, $\delta_\mathrm{J2000}$) $\sim$ ($05^\mathrm{h}08^\mathrm{m}58\fs0$, $-68\degr43\arcmin28\arcsec$), $\sim$ ($05^\mathrm{h}08^\mathrm{m}58\fs0$, $-68\degr43\arcmin35\arcsec$), and $\sim$ ($05^\mathrm{h}08^\mathrm{m}58\fs3$, $-68\degr43\arcmin33\arcsec$) (see dashed circles in Figure \ref{fig1}a). These X-ray spots are also bright in the H$\alpha$ emission. Figure \ref{fig1}b shows an H$\alpha$ image obtained using {\it{Hubble Space Telescope}} (HST) with WFC3. We see prominent optical nebular knots (I), (II), (III), and (IV) defined as \citep[][dashed circles in Figure \ref{fig1}b]{2017ApJ...836...85L}. The diffuse H$\alpha$ emission is seen in only the western half of the SNR, similar to the soft-band X-ray distribution.

\begin{figure}
\begin{center}
\includegraphics[width=90mm,clip]{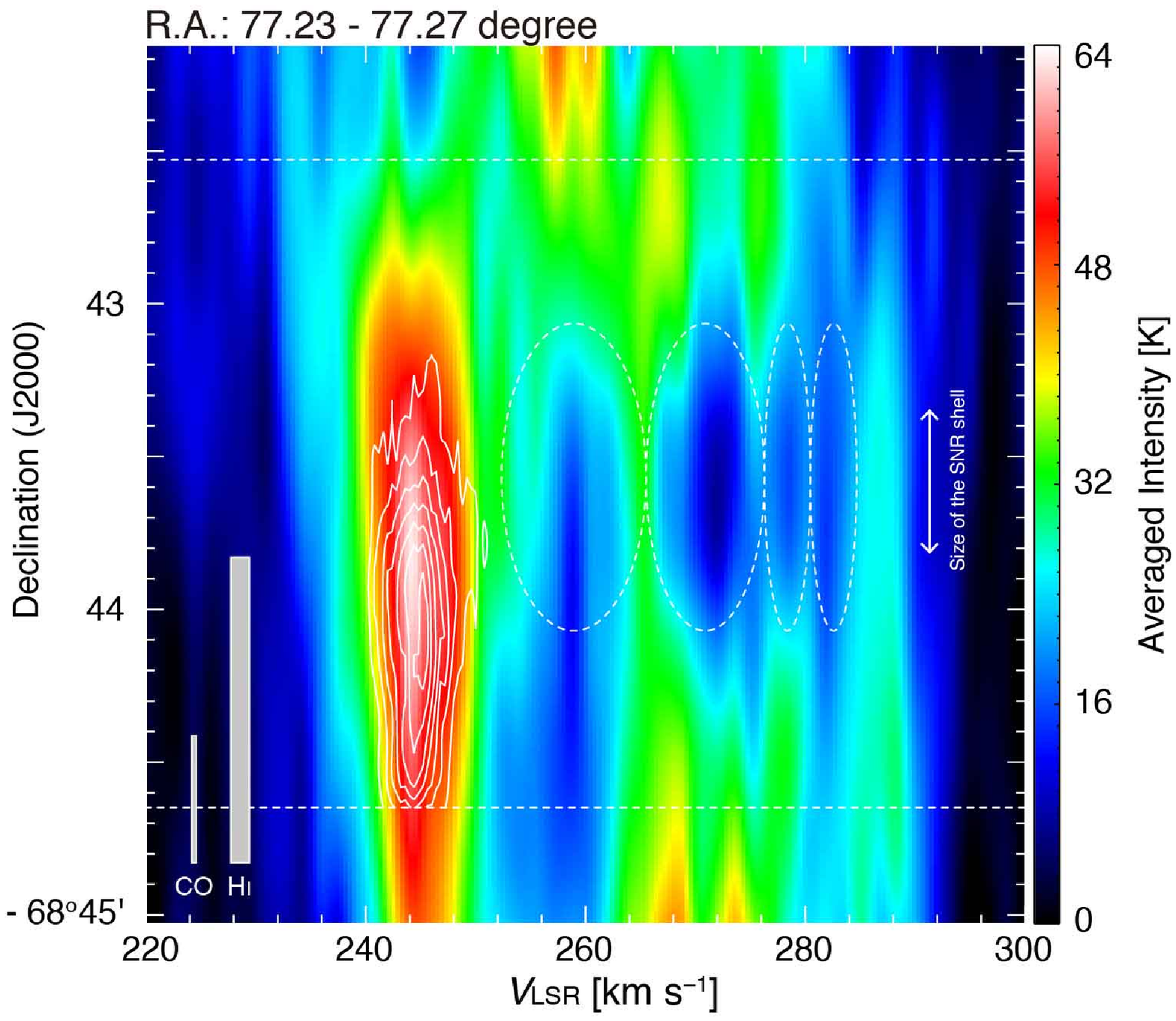}
\caption{Position--velocity diagram of ATCA $\&$ Parkes H{\sc i} ({\it{color image}}) and ASTE $^{12}$CO($J$ = 3--2) ({\it{contours}}). The integration velocity range in Right Ascension is from 77.23 to 77.27 degrees, corresponding to the HPBW of H{\sc i} data ($\sim60''$). The lowest contour level and contour intervals are 0.28 K ($\sim7\sigma$). The beam size and velocity resolution are also shown. Horizontal lines represent observed boundaries of CO. Dashed circles indicate the H{\sc i} hole-like structures in the velocity space. The vertical arrow indicates the size of the SNR shell.}
\label{fig3}
\end{center}
\end{figure}%

\begin{figure}
\begin{center}
\includegraphics[width=90mm,clip]{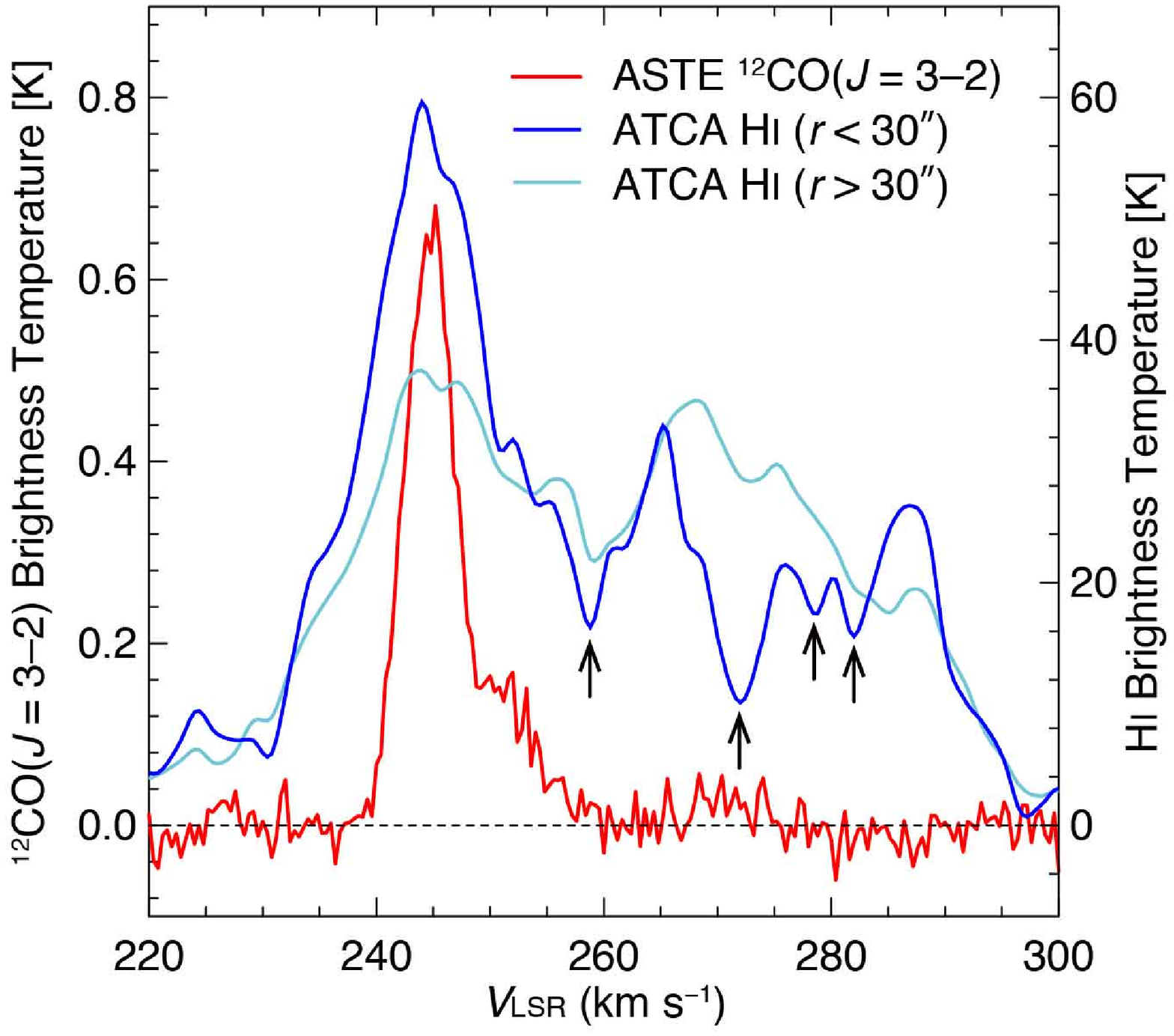}
\caption{Averaged profiles of $^{12}$CO($J$ = 3--2) ({\it{red}}) and H{\sc i} ({\it{blue}} and {\it{cyan}}) toward the SNR N103B. The regions averaged the cyan and blue spectra are the outside and inside of the circle with a radius of $30''$ at the center position of ($\alpha_\mathrm{J2000}$, $\delta_\mathrm{J2000}$) $\sim$ ($05^\mathrm{h}08^\mathrm{m}59^\mathrm{s}$, $-68\degr43\arcmin36\arcsec$). The boundaries averaged both the CO and H{\sc i} spectra are the map size of Figure \ref{fig2}.}
\label{fig4}
\end{center}
\end{figure}%

Figure \ref{fig2} shows the velocity channel maps of $^{12}$CO($J$ = 3--2) and H{\sc i} obtained using ASTE and ATCA \& Parkes, respectively. We confirmed an existence of GMC in the south of the SNR, previously mentioned by \cite{2017AIPC.1792d0038S,sano2018}. The size of GMC is $\sim$30 pc $\times$ 20 pc with an arc-like morphology. The southern edge of the SNR appears to be along the GMC in the velocity range of 236--252 km s$^{-1}$ (see Figure \ref{fig2}a). We newly find an HI cloud in the same velocity of the GMC (Figure \ref{fig2}d). To derive the mass of the H{\sc i} cloud, we use the following equation \citep{1990ARA&A..28..215D}:
\begin{eqnarray}
N_\mathrm{p}(\mathrm{H}{\textsc{i}}) = 1.823 \times 10^{18} \; W(\mathrm{H}{\textsc{i}}) \;(\mathrm{cm}^{-2}),  
\label{eq1}
\end{eqnarray}
where $N_p$(H{\sc i}) is the proton column density of H{\sc i} and $W$(H{\sc i}) is the integrated intensity of H{\sc i}. We finally obtain the mass of the H{\sc i} cloud is $\sim$$2 \times 10^4$ $M_\sun$, which is significantly smaller than that of the GMC ($\sim$$10^6$ $M_\sun$). We also note that the diffuse H{\sc i} gas is also seen in the velocity of $V_\mathrm{LSR} > 250$ km s$^{-1}$. For the H{\sc i} map of the velocity of 268--284 km s$^{-1}$, the H{\sc i} integrated intensity is decreased toward the SNR (see Figure \ref{fig2}f). The size of intensity-decreased region coincides with the beam size of H{\sc i} data.

Figure \ref{fig3} shows position-velocity diagrams of H{\sc i} and CO. The peak velocity of GMC is quite similar to that of H{\sc i} cloud ($V_\mathrm{LSR}\sim$244 km s$^{-1}$). We also find four cavity-like structures of H{\sc i} in the position-velocity diagram (see dashed circles in Figure \ref{fig3}). The cavity-like structures are centered at the position of the SNR in declination. The sizes of them are similar to the beam size of H{\sc i}. This indicates that the H{\sc i} gas at the velocity of $V_\mathrm{LSR} > 250$ km s$^{-1}$ is observed as the H{\sc i} absorption line owing to the strong radio continuum emission from the SNR, and hence is located in front of the SNR. A similar case is also seen in the LMC SNRs N49 \citep[][]{2018ApJ...863...55Y}. The H{\sc i} profiles also support this idea. Figure \ref{fig4} shows the CO and H{\sc i} profiles toward the SNR N103B. We find significant differences between the ambient H{\sc i} gas (cyan, outside of the SNR) and the H{\sc i} gas in the direction of the SNR (blue, inside of the SNR). The velocities of absorption lines indicated by the four arrows roughly coincide with the central velocities of the H{\sc i} cavity-like structures as shown in Figure \ref{fig3}. We also find the line broadening of CO at the velocity range from $\sim$248 to $\sim$260 km s$^{-1}$, implying a possible evidence for the acceleration of part of the GMC. In the present paper, we therefore focus on the GMC and H{\sc i} cloud at the velocity around 250 km s$^{-1}$, which are likely associated with the SNR.

\begin{figure*}
\begin{center}
\includegraphics[width=\linewidth,clip]{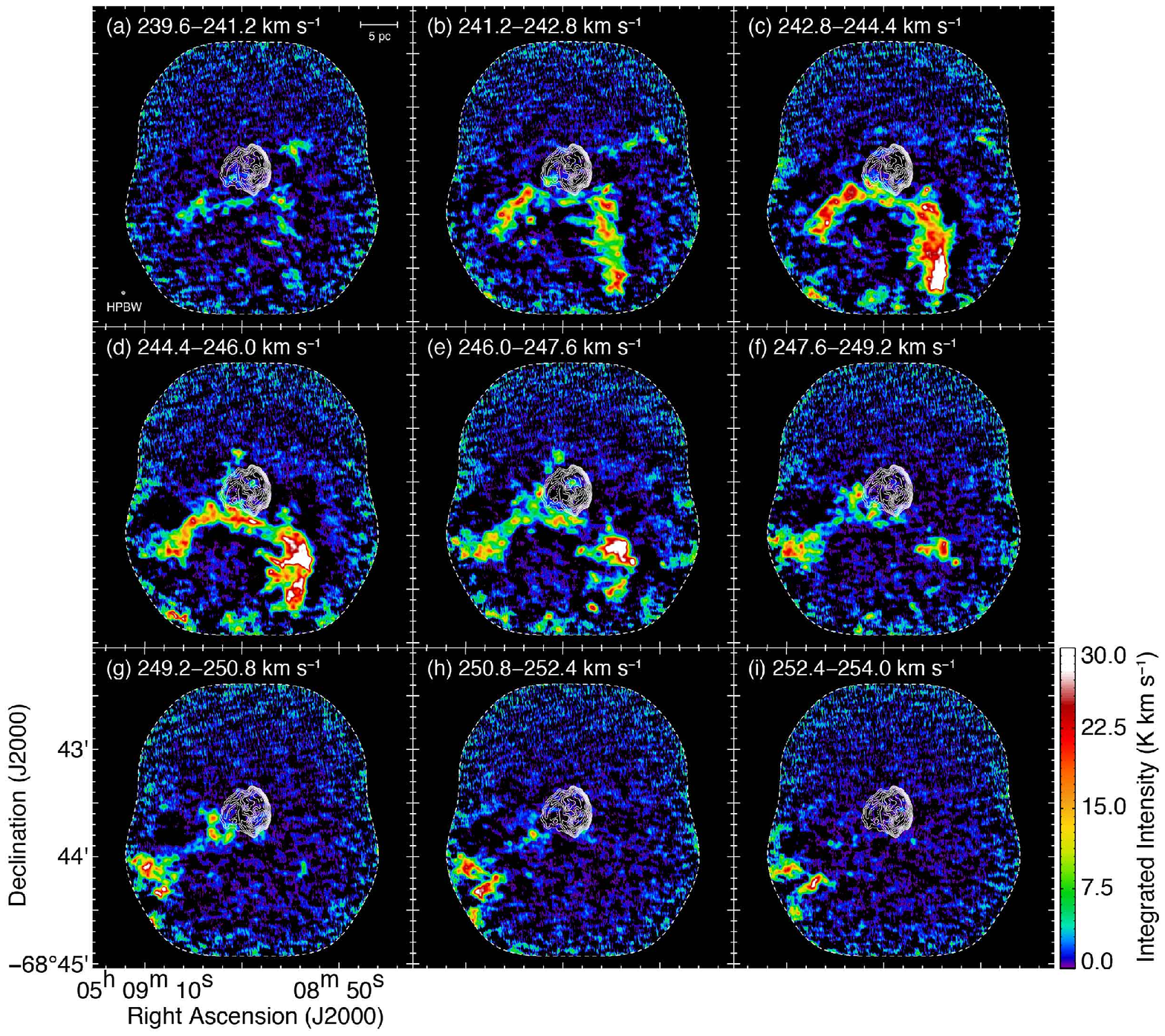}
\caption{Velocity channel map of the ALMA $^{12}$CO($J$ = 1--0) superposed on the {\it{Chandra}} X-ray contours as in Figure \ref{fig2}. Each panel shows CO intensity distribution integrated every 1.6 km s$^{-1}$ in a velocity range from 239.6 to 254.0 km s$^{-1}$. The regions enclose by the dashed lines indicate the observed areas using ALMA.}
\label{fig5}
\end{center}
\end{figure*}%

\subsection{Detailed CO Distribution with ALMA}\label{subsec:alma}
Figure \ref{fig5} shows velocity channel maps of $^{12}$CO($J$ = 1--0) line emission obtained with ALMA, where the {\it{Chandra}} X-ray contours are superposed. Each map taken every 1.6 km s$^{-1}$ from $V_\mathrm{LSR}$ = 239.6--254.0 km s$^{-1}$. The GMC in N103B is spatially resolved into many CO filaments and clouds, and only part of the GMC is associated with the SNR. For the velocity range of $V_\mathrm{LSR}$ = 244.4--246.0 km s$^{-1}$ (Figure \ref{fig5}d), CO filaments and the edge of the GMC completely delineate the outer boundary of the SNR from south to northeast in a clockwise direction. This trend is also seen at the velocity range of $V_\mathrm{LSR}$ = 246.0--252.4 km s$^{-1}$ (Figures \ref{fig5}e--\ref{fig5}h). We note that tiny molecular clumps at ($\alpha_\mathrm{J2000}$, $\delta_\mathrm{J2000}$) $\sim$ ($05^\mathrm{h}08^\mathrm{m}58\fs9$, $-68\degr43\arcmin31\arcsec$) (hereafter referred to as ``clump A''), and at ($\alpha_\mathrm{J2000}$, $\delta_\mathrm{J2000}$) $\sim$ ($05^\mathrm{h}08^\mathrm{m}58\fs1$, $-68\degr43\arcmin46\arcsec$) (hereafter referred to as ``clump B''). The clump A appears to be embedded within the SNR, while the clump B is associated with the edge of the southwestern shell.

\begin{figure}
\begin{center}
\includegraphics[width=\linewidth,clip]{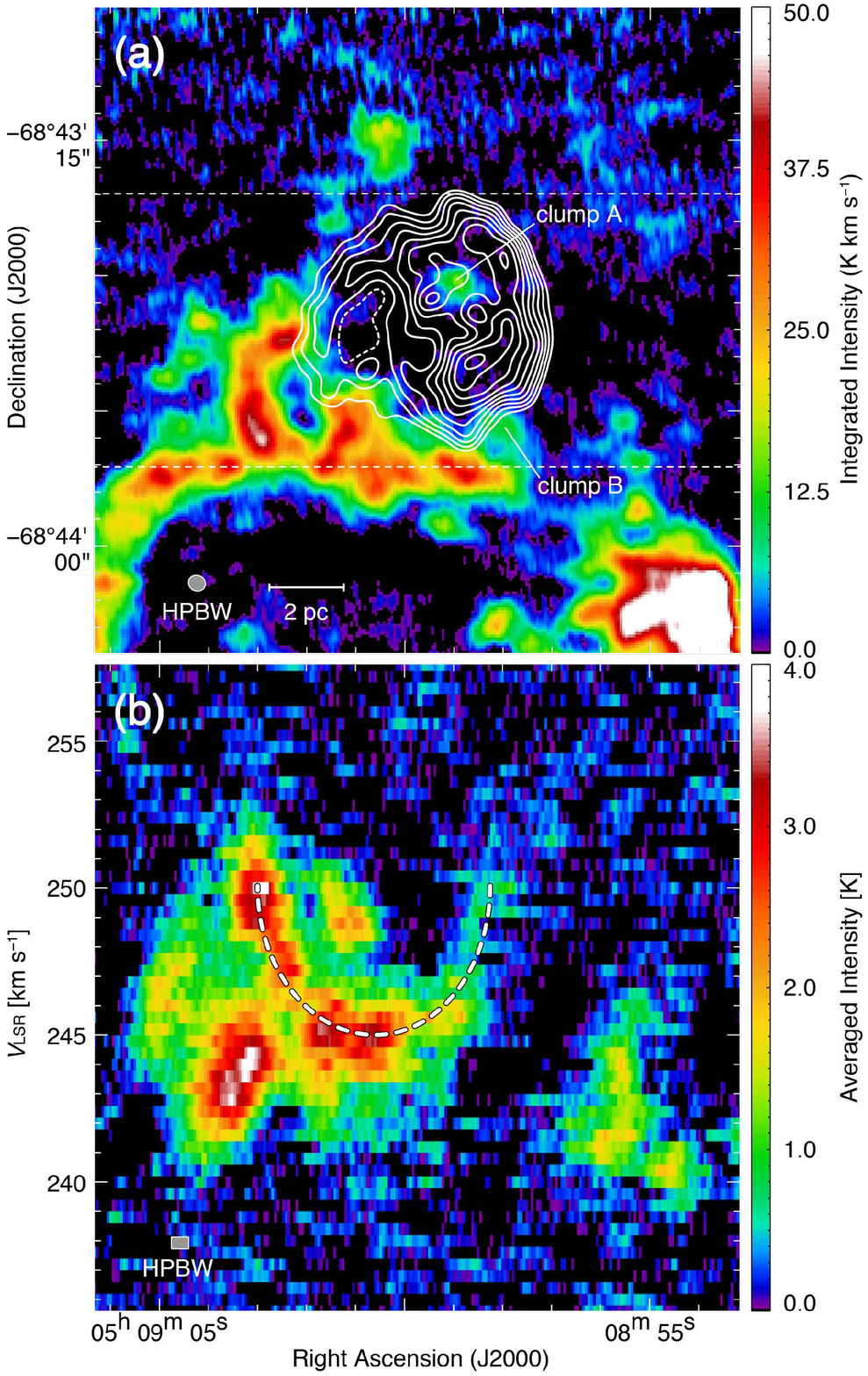}
\caption{(a) Integrated intensity map of ALMA $^{12}$CO($J$ = 1--0) superposed on the {\it{Chandra}} X-ray contours. The integration velocity range is from 244.8 to 252.8 km s$^{-1}$. The contour levels are the same as in Figure \ref{fig2}. The tiny molecular clumps A and B are also shown. (b) Position--velocity diagram of the ALMA $^{12}$CO($J$ = 1--0). The integration range in Declination is from $-68\degr43\arcmin51\arcsec$ to $-68\degr43\arcmin21\arcsec$, corresponding to the dashed lines in Figure \ref{fig6}a. The dashed-curve represents an expanding gas motion.}
\label{fig6}
\end{center}
\end{figure}%

Figure \ref{fig6}a shows an enlarged view of the $^{12}$CO($J$ = 1--0) distribution in a velocity range from $V_\mathrm{LSR}$ = 244.8--252.8 km s$^{-1}$. The molecular clouds show a remarkably good spatial correlation with the SNR shell from southwest to northeast in clockwise direction, indicating that the molecular clouds are likely associated with the SNR shockwaves. By contrast, no molecular clouds are detected in the northwestern shell except for the tiny molecular clump embedded within the shell. Figure \ref{fig6}b shows a position-velocity diagram of $^{12}$CO($J$ = 1--0) with the integration range from $-68\degr43\arcmin51\arcsec$ to $-68\degr43\arcmin21\arcsec$ in Declination (see dashed lines in Figure \ref{fig6}a). We find a cavity-like structure in the position-velocity diagram, indicating an evidence for the expanding gas motion with an expansion velocity of $\sim$5 km s$^{-1}$ (see the dashed-curve in Figure \ref{fig6}b). The spatial extent of cavity-like structure is roughly similar to the shell size of the SNR.

To estimate the mass of molecular clouds, we use the following equation:
\begin{eqnarray}
M = \mu m_{\mathrm{p}} D^2 \Omega \sum_{i} [N_i(\mathrm{H}_2)],
\label{eq2}
\end{eqnarray}
where $\mu$ is the mean molecular weight relative to the atomic hydrogen, $m_\mathrm{p}$ is the mass of atomic hydrogen, $D$ is the distance to the molecular cloud, $\Omega$ is the solid angle of a pixel, and $N_i(\mathrm{H}_2)$ is the molecular hydrogen column density for each pixel $i$. We use $\mu = 2.38$ by taking into account the helium abundance $\sim$36\% to the molecular hydrogen mass \citep[c.f.,][]{2008ApJS..178...56F}. The molecular hydrogen column density $N(\mathrm{H}_2)$ is given by using the relation: 
\begin{eqnarray}
N(\mathrm{H}_2) = X \cdot W(^{12}\mathrm{CO}),
\label{eq3}
\end{eqnarray}
where $X$ is the conversion factor in the unit of cm$^{-2}$ (K km s$^{-1})^{-1}$ and $W(^{12}\mathrm{CO})$ is the integrated intensity of the $^{12}$CO($J$ = 1--0) line emission in the unit of K km s$^{-1}$. We use the conversion factor $X = 7.0 \times 10^{20}$ cm$^{-2}$ (K km s$^{-1}$)$^{-1}$ \citep{2008ApJS..178...56F}. The physical properties of molecular clouds are estimated for the regions that are significantly detected by CO line emission with a 3-sigma or higher. The size of molecular cloud is defined as $(A / \pi)^{0.5} \times 2$, where $A$ is the total surface area of the molecular cloud enclosed by the 3-sigma level. We finally derive the total mass of the molecular cloud of $\sim$$8 \times 10^5$ $M_\sun$, which is roughly consistent with previous CO studies \citep{2017AIPC.1792d0038S,sano2018}.

\begin{figure*}
\begin{center}
\includegraphics[width=\linewidth,clip]{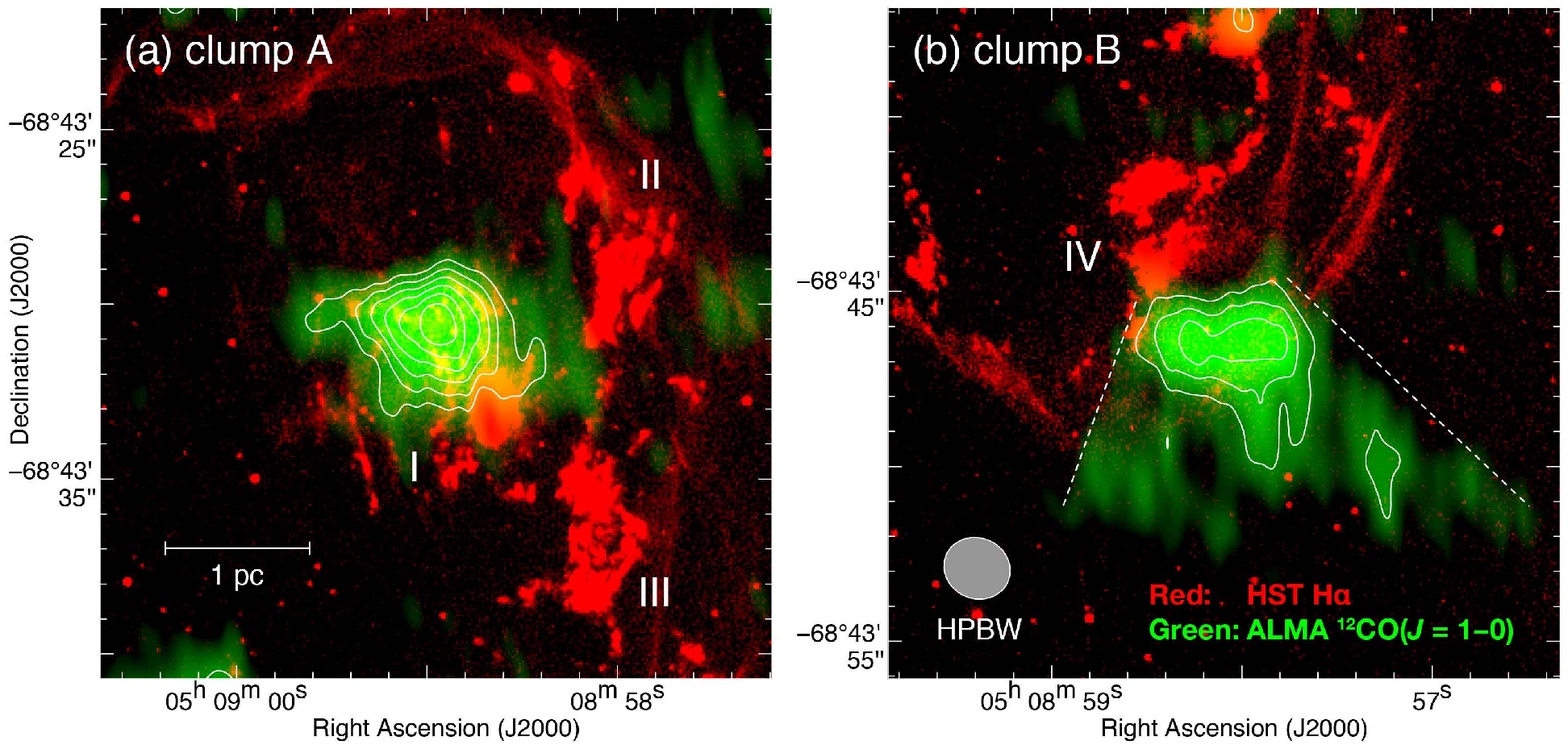}
\caption{Maps of HST H$\alpha$ intensity ({\it{red color}}) and integrated intensity of the ALMA $^{12}$CO($J$ = 1--0) ({\it{green color}}) toward (a) clump A and (b) clump B. The integration velocity range is 244.8 to 248.0 km s$^{-1}$ for clump A and is 249.2 to 252.0 km s$^{-1}$ for clump B. Superposed contours indicate the CO intensity and are plotted from 4 K km s$^{-1}$ with 2 K km s$^{-1}$ steps. We also showed the optical nebular knots (I), (II), (III), and (IV) as defined by \cite{2017ApJ...836...85L}.}
\label{fig7}
\end{center}
%\vspace*{-1cm}
\end{figure*}%

\subsection{Tiny Molecular Clumps Associated with the Optical Nebula Knots}\label{subsec:tiny}
Figure \ref{fig7} shows enlarged views toward tiny molecular clumps A and B. Red and green images represent the HST H$\alpha$ and ALMA $^{12}$CO($J$ = 1--0) integrated intensity, respectively. We find that the tiny molecular clumps A and B are located near the optical nebular knots (I) and (II), which are also bright in the soft-band X-rays, respectively (see Figure \ref{fig1}a). In contrast, there are no CO counterparts for other optical nebular knots (II) and (III) in Figure \ref{fig7}a. Since the tiny clumps have extended distributions compared with the beam size of ALMA, we assumed a beam-filling factor of $\sim$1. Using the equations (2) and (3), the size, mass, and density are therefore estimated to be $\sim$110 $M_\sun$, $\sim$1.2 pc, and $\sim$2000 cm$^{-3}$ for clump A; $\sim$90 $M_\sun$, $\sim$1.3 pc, and $\sim$1400 cm$^{-3}$ for clump B. We also note that the molecular clump B shows a V-shaped structure pointing toward the center of the SNR (see dashed lines in Figure \ref{fig7}b), which will be discussed later. 

\section{Discussion}\label{sec:dis}
\subsection{Inhomogeneous Distribution of the Interstellar Gas}\label{subsec:inhomo}
The dynamical interaction between the SNR shocks and the ISM affects the evolution of SNRs through a strong deformation of the shell morphology, if the ISM is dense enough (e.g., RX~J1713.7$-$3946, \citeauthor{2003PASJ...55L..61F} \citeyear{2003PASJ...55L..61F}; Kes~79, \citeauthor{2017arXiv171108165K} \citeyear{2017arXiv171108165K}). On the other hand, Type Ia SNRs usually evolve in the low-density ambient medium, and hence form a relatively symmetric morphology \citep[][]{2009ApJ...706L.106L}. According to \cite{2018MNRAS.tmp.1139R}, the LMC Type Ia SNR J0509$-$6731 shows a symmetry shell with an isotropic expansion velocity of $\sim$$7500 \pm 1700$ km s$^{-1}$ derived by the proper motion studies in X-rays. No dense molecular cloud is detected toward the SNR and the column density of H{\sc i} is estimated to be $\sim$4--$6 \times 10^{20}$ cm$^{-2}$, indicating a very low-density interstellar environment. In this section, we discuss the interstellar environment in the Type Ia SNR N103B and its relation with the shell morphology.

Based on the ALMA results, the CO clouds---a part of the GMC---are physically associated with the SNR N103B. The CO clouds delineate the outer boundary of the SNR from southwest to northeast in clockwise direction, while the northwestern shell has no CO clouds (see Figures \ref{fig5} and \ref{fig6}a). Although the spatial resolution is limited of H{\sc i} data ($\sim$15 pc), we estimate the peak column density of H{\sc i} is $\sim$$2 \times10^{21}$ cm$^{-2}$. We argue that the inhomogeneous CO distribution and gas rich environment possibly affect the X-ray shell-morphology of the SNR. In fact, the northwestern edge of X-ray shell shows a spherical shape, whereas the southeastern shell appears to be strongly deformed (see Figure \ref{fig1}a). Moreover, a difference of shock velocity for each region has been observed. According to \cite{2017ApJ...847..122G}, the velocity for Balmer-dominant shocks is $\sim$2100 km s$^{-1}$ in the northern edge of the shell, in which no molecular clouds are seen. On the other hand, the slow shock velocity of $\sim$1200 km s$^{-1}$ is observed in the southwestern edge of the SNR where the tiny molecular clump B is embedded  (see Figures \ref{fig6}a and \ref{fig7}b). Invoking ram pressure conservation ($\rho V_\mathrm{sh}^2$ = constant), we expect that the ISM density of the southeastern shell is at least three times higher than that of the northern shell. To verify the interpretation, further H{\sc i} observations with high spatial resolution with a few pc scales are needed. In addition, direct measurements of shock velocity using the proper motion technique is useful to compare the inhomogeneity of the ISM associated with the SNR.

\begin{figure*}
\begin{center}
\includegraphics[width=\linewidth,clip]{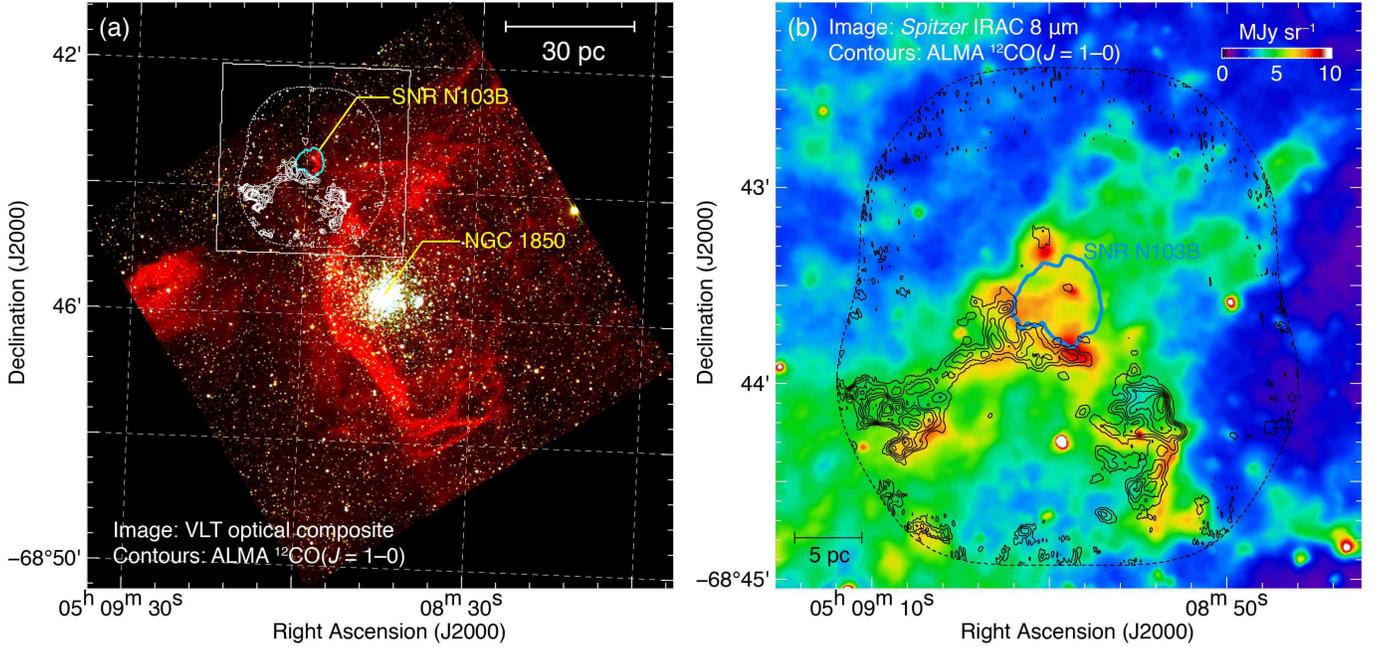}
\caption{(a) Optical image of NGC~1850 taken with the Very Large Telescope (VLT). The red, green, and blue represent H$\alpha$, V-, and, B-bands respectively (credit: ESO). White contours indicate the ALMA $^{12}$CO($J$ = 1--0) integrated intensity ($V_\mathrm{LSR}$ = 244.8--252.8 km s$^{-1}$) and the lowest contour level and contour intervals are 10 K km s$^{-1}$. The dashed contour represents the {\it{Chandra}} X-ray boundary at the level of $0.2 \times 10^{-5}$ photons s$^{-1}$ pixel$^{-1}$. The rectangle box indicates the region shown in Figure \ref{fig8}b. (b) SAGE {\it{Spitzer}} IRAC 8 $\mu$m image toward the SNR N103B \citep{2006AJ....132.2268M}. The superposed contours are the same as in Figure \ref{fig8}a.}
\label{fig8}
\end{center}
%\vspace*{-3cm}
\end{figure*}

Moreover, the origin of inhomogeneous gas distribution is possibly related with the stellar environment. In fact, the SNR is projected near the double star cluster NGC~1850, which contains a young ($50 \pm 10$ Myr) globular-like star cluster and an even younger ($4.3 \pm 0.9$ Myr) small cluster \citep[e.g.,][]{1994ApJ...435L..43G}. Figure \ref{fig8}a shows the optical image around the SNR N103B obtained with the Very Large Telescope (VLT). The globular-like star cluster NGC~1850 is located toward the southwest of N103B with a projected separation between them of $\sim$30 pc, indicating that the molecular clouds surrounding N103B are affected by the stellar feedback from the star clusters. \cite{2014PASJ...66...26S} also suggested the physical association of N103B and NGC~1850 by X-ray spectroscopic analysis. On the other hand, \cite{2016A&A...585A.162M} argued that the spatial coincidence between the H$\alpha$ nebula and the SNR is likely a projection effect. They concluded that the SNR is located on the far side of the LMC as suggested by its large fraction of $N_\mathrm{p}$(X-rays) / $N_\mathrm{p}$(H{\sc i}), where $N_\mathrm{p}$(X-rays) is the absorbing column density of N103B. However, the fraction is underestimated because they did not take into account the molecular hydrogen as the ISM component. We claim that the SNR N103B is located on the similar distance of the globular-like star cluster NGC 1850. Figure \ref{fig8}b shows the {\it{Spitzer}} IRAC 8 $\mu$m image toward N103B superposed on the CO contours. We find that the bright spots in 8 $\mu$m---yellow or red colors in Figure \ref{fig8}b---are located only the southwestern edges of molecular clouds except for the direction of the SNR. Since the {\it{Spitzer}} IRAC 8 $\mu$m image is dominated by polycyclic aromatic hydrocarbon (PAH) line emission, this trend may be understood with the UV excitation by the massive star clusters. Future ALMA and SOFIA observations with [C{\sc i}] and [C{\sc ii}] will allow us to study physical properties of PDR under the strong feedback in addition to the origin of the inhomogeneous gas distribution.

\subsection{Implications for the progenitor system of N103B}\label{subsec:progenitor}
SNR N103B is one of the most promising candidates for the SD hypothesis, that is, for having a progenitor system consisting of a white dwarf and a non-degenerate companion \citep[e.g.,][]{2014ApJ...790..139W,2017ApJ...836...85L,2017ApJ...847..122G}. The SD progenitor system may produce optically thick winds---named as accretion winds---from the white dwarf when the mass accretion rate is higher than the stable nuclear fusion rate \citep[e.g.,][]{1996ApJ...470L..97H,1999ApJ...522..487H,1999ApJ...519..314H}. If this is true, the expanding gas motion of the ISM can be observed similarly to the other Type Ia SNRs: RCW~86 and Tycho \citep{2017JHEAp..15....1S,2017A&A...604A..13C}. We here discuss the progenitor system of N103B based on the ALMA CO results of the associated interstellar molecular gas.

First, we argue that the expanding shell of CO cloud (Figure \ref{fig6}b) is roughly consistent with the acceleration by only the accretion winds of the progenitor system of N103B. Based on the ALMA CO results and the equations (\ref{eq2}) and (\ref{eq3}), the mass of expanding shell is estimated to be $\sim$$5 \times 10^3$ $M_\sun$. Since the expansion velocity of CO is $\sim$5 km s$^{-1}$, the momentum is calculated to be $\sim$$2.5 \times 10^4$ $M_\sun$ km s$^{-1}$. According to \cite{1999ApJ...522..487H,1999ApJ...519..314H}, the accretion wind of Type Ia SD progenitor system has a wind mass of $\sim$ $10^{-4}$ $M_\sun$ yr$^{-1}$ and a wind velocity of $\sim$2000 km s$^{-1}$ as maximum values \citep[see also][]{2005ASPC..342..105N}. Assuming the wind duration of $\sim$ $3 \times 10^5$ yr, the momentum of total accretion wind is to be $\sim$$6 \times 10^4$ $M_\sun$ km s$^{-1}$. It is therefore that $\sim$40\% of the wind momentum had transferred to the expanding gas motion. The percentage of momentum transfer is slightly larger than a theoretical calculation in an ideal adiabatic wind shell \citep[e.g.,][]{1977ApJ...218..377W}, but is not impossible. In addition to this, the CO cloud is possibly accelerated by the shock wave of N103B even though the young age of 380--860 yr \citep{2005Natur.438.1132R}. According to \cite{2016ApJ...826...34Z}, a possible line broadening of CO emission is detected in the young SNR Tycho with an age of 446 yr. To test the gas acceleration and line broadening in N103B, we need more sensitive CO observations with fine spatial resolutions with ALMA.

We also note that the V-shaped structure of tiny molecular clump B is possibly evidence for a blown-off cloud that has overtaken the accretion winds from the progenitor system. According to \cite{2017ApJ...850...71F}, a similar V-shaped structure of H{\sc i}/CO clouds is observed in the Galactic core-collapse SNR Vela~Jr. The authors conclude that a V-shaped H{\sc i} tail with a dense head of CO survived against the stellar wind from the massive progenitor. Since the accretion wind of N103B is weaker than the stellar wind of Vela~Jr, the V-shaped structure of N103B was possibly formed by the accretion wind because of the high wind speed up to $\sim$2000 km s$^{-1}$ and the large wind duration.

In conclusion, we propose a possible scenario that N103B exploded in the wind-bubble formed by the accretion winds from the progenitor system, and is now interacting with the dense wind-cavity of the molecular clouds. This scenario is consistent with the SD origin of N103B. To confirm the scenario, a reliable identification of companion a star is needed.

\section{Conclusions}\label{sec:conc}
We have carried out new $^{12}$CO($J$ = 1--0, 3--2) observations using ALMA and ASTE in order to spatially resolve molecular clouds associated with the Type Ia SNR N103B in the LMC. The primary conclusions of the present study are summarized as below.

\begin{enumerate}
\item We have confirmed the existence of GMC at $V_\mathrm{LSR}$ $\sim$245 km s$^{-1}$ toward the southeast of the SNR using the $^{12}$CO($J$ = 3--2) data at the spatial resolution of $\sim$6 pc. Using the ALMA $^{12}$CO($J$ = 1--0) data, we have spatially resolved CO clouds along the southeastern shell of the SNR with the spatial resolution of $\sim$0.4 pc. The CO clouds show the expanding gas motion in the position-velocity diagram with the expansion velocity of $\sim$5 km s$^{-1}$. The spatial extent of expanding shell is roughly similar to the size of the SNR. On the other hand, there are no molecular clouds toward the northwestern shell of the SNR. We also revealed tiny molecular clumps associated with the optical nebula knots, the size and mass of which are estimated to be $\sim$100 $M_\sun$ and $\sim$1 pc, respectively.
\item We have argued that the asymmetric morphology of the X-ray shell is possibly related to the inhomogeneous distribution of the ISM. For the northwestern region, the SNR has the spherical shell in X-rays because there are no dense molecular clouds. On the other hand, the southeastern shell is strongly deformed owing to the shock interaction with the dense molecular clouds. The shock-velocities derived by optical emission are also consistent with the inhomogeneous density distribution of the ISM. To verify the interpretation, both the further H{\sc i} observations with fine spatial resolution and the direct measurements of shock velocity---e.g., proper motion measurements---are needed.
\item We have presented a possible scenario that N103B exploded in the wind-bubble formed by the accretion winds from the progenitor system, and is now interacting with the wind-cavity of the molecular gas. Because the momentum of the expanding gas motion can be explained by the energy input of the accretion wind from the SD progenitor system. To confirm the scenario, a reliable identification of companion a star is needed.
\end{enumerate}

\acknowledgments
We are grateful to Aya Bamba, Ping Zhou, and Satoru Katsuda for thoughtful comments on the X-ray properties of N103B and the CSM or wind-bubble in Type Ia SNRs. This paper makes use of the following ALMA data: ADS/JAO.ALMA$\#$2015.1.01130.S. ALMA is a partnership of ESO (representing its member states), NSF (USA) and NINS (Japan), together with NRC (Canada) and NSC and ASIAA (Taiwan) and KASI (Republic of Korea), in cooperation with the Republic of Chile. The Joint ALMA Observatory is operated by ESO, AUI/NRAO and NAOJ. The scientific results reported in this article are based on data obtained from the Chandra Data Archive (Obs IDs: 125 and 3810). This research has made use of software provided by the Chandra X-ray Center (CXC) in the application packages CIAO (v 4.10). The Australia Telescope Compact Array and the Parkes radio telescope are part of the ATNF which is funded by the Australian Government for operation as a National Facility managed by CSIRO. This study was financially supported by Grants-in-Aid for Scientific Research (KAKENHI) of the Japanese Society for the Promotion of Science (JSPS, grant Nos. 15H05694 and 16K17664). HS was supported by ``Building of Consortia for the Development of Human Resources in Science and Technology'' of Ministry of Education, Culture, Sports, Science and Technology (MEXT, grant No. 01-M1-0305). K. Tokuda was supported by NAOJ ALMA Scientific Research Grant Number of 2016-03B. We are grateful to the anonymous referee for useful comments, which helped the authors to improve the paper.
\software{CASA \citep[v 5.1.0.:][]{2007ASPC..376..127M}, CIAO \citep[v 4.10:][]{2006SPIE.6270E..1VF}}

\end{document}